\documentclass[twocolumn,prx,amsmath,amssymb,superscriptaddress,longbibliography]{revtex4-1}

\usepackage{lipsum}
\usepackage[T1]{fontenc}
\usepackage{bbold}
\usepackage{amsthm}

\usepackage{dsfont}
\usepackage{soul}
\usepackage{bm}
\usepackage{color} 
\usepackage{comment}
\usepackage{mathtools}
\usepackage{float}
\usepackage{xcolor}
\usepackage{physics, calc} 
\usepackage{graphicx}
\usepackage{enumitem}
\usepackage[ruled,vlined]{algorithm2e}
\usepackage[citecolor=blue, urlcolor=blue, linkcolor=blue, colorlinks=true]{hyperref}
\usepackage{dirtytalk}
\usepackage{dsfont}

\DeclareMathOperator*{\argmax}{arg\,max}
\makeatletter
\newcommand*\rel@kern[1]{\kern#1\dimexpr\macc@kerna}
\newcommand*\widebar[1]{%
  \begingroup
  \def\mathaccent##1##2{%
    \rel@kern{0.8}%
    \overline{\rel@kern{-0.8}\macc@nucleus\rel@kern{0.2}}%
    \rel@kern{-0.2}%
  }%
  \macc@depth\@ne
  \let\math@bgroup\@empty \let\math@egroup\macc@set@skewchar
  \mathsurround\z@ \frozen@everymath{\mathgroup\macc@group\relax}%
  \macc@set@skewchar\relax
  \let\mathaccentV\macc@nested@a
  \macc@nested@a\relax111{#1}%
  \endgroup
}

\makeatother

\definecolor{azure}{rgb}{0.0, 0.5, 1.0}

\newcommand{\TUM}{\affiliation{Department of Physics, TFK, Technische Universit{\"a}t M{\"u}nchen, James-Franck-Stra{\ss}e 1, D-85748 Garching, Germany}}
\newcommand{\MCQST}{\affiliation{Munich Center for Quantum Science and Technology (MCQST), Schellingstr. 4, 80799 M{\"u}nchen, Germany}}

\begin{document}
\raggedbottom

\title{Data compression for quantum machine learning}
\author{Rohit Dilip}
\TUM
\affiliation{Division of Chemistry and Chemical Engineering, California Institute of Technology, Pasadena, California 91125, USA}
\author{Yu-Jie Liu}
\TUM\MCQST
\author{Adam Smith}
\affiliation{School of Physics and Astronomy, University of Nottingham, Nottingham, NG7 2RD, UK}
\affiliation{Centre for the Mathematics and Theoretical Physics of Quantum Non-Equilibrium Systems, University of Nottingham, Nottingham, NG7 2RD, UK}
\author{Frank Pollmann}
\TUM\MCQST

\begin{abstract}The advent of noisy-intermediate scale quantum computers has introduced the exciting possibility of achieving quantum speedups in machine learning tasks. These devices, however, are composed of a small number of qubits, and can faithfully run only short circuits. This puts many proposed approaches for quantum machine learning beyond currently available devices. We address the problem of compressing classical data into efficient representations on quantum devices.
Our proposed methods allow both the required number of qubits and depth of the quantum circuit to be tuned. We achieve this by using a correspondence between matrix-product states and quantum circuits, and further propose a hardware-efficient quantum circuit approach, which we benchmark on the Fashion-MNIST dataset. Finally, we demonstrate that a quantum circuit based classifier can achieve competitive accuracy with current tensor learning methods using only 11 qubits.
\end{abstract}

\maketitle
\section{Introduction}
The rapid development of quantum computers has spurred proposals for quantum speedups in many fields, not least for applications in machine learning. One direction can be summarized as quantum-enhanced machine learning, where quantum algorithms are applied to classical data ~\cite{biamonte2017quantum}. Exploration in this direction has led to various quantum machine learning algorithms~\cite{bokhan2022multiclass, sasaki2002,rebentrust2014,lu2014,lloyd2013,Havlicek2019}. In certain settings, the use of fault-tolerant quantum computers provides a provable advantage over classical approaches~\cite{Liu2021,Huang2021}. However, the significant resource cost of these methods makes finding practical methods for noisy intermediate-scale quantum (NISQ) devices a significant priority.

A recurring problem is the loading of classical data into quantum machine learning algorithms. A typical approach is to represent the data as a quantum state, which can be done in multiple ways. For example, one could encode a black and white image by mapping the classical bits with value 0 and 1 to the corresponding qubit (quantum bit) states of the quantum computer~\cite{Schuld2018,Farhi2018}. While conceptually simple and easy to implement with a single layer quantum circuit, even modestly sized images with hundreds of pixels--such as Fashion-MNIST~\cite{xiao2017fashion}--would be well beyond the qubit capabilities of current devices. By leveraging quantum entangling operations, it is also possible to encode the image in a state of a logarithmic number of qubits~\cite{Le2010,latorre2005}. For the same example of Fashion-MNIST, this requires a much more manageable 10 qubits, but the resulting circuit is too deep for the current fidelity of gates and qubit coherence times.

In parallel, tensor network methods have been applied to basic data compression and machine learning tasks with near state-of-the-art performance~\cite{stoudenmire2016supervised, efthymiou2019tensornetwork, huggins2019,bengua2017}. These methods represent quantum states as a product of tensors with cutoff parameters -- bond dimensions -- that allow for systematic approximations of the quantum state by limiting the quantum entanglement. Importantly, a large class of these tensor networks known as matrix-product states (MPS)~\cite{fannes1992finitely, Schollwock2011}, can be directly mapped to quantum circuits with a depth that scales polynomially in the number of qubits and the bond dimension~\cite{Schon2005a, Smith2019b, ran2020encoding, Barratt2020, lin2021real}. This contrasts with the exponential scaling with number of qubits for exact parameterizations of generic states.

In this work, we resolve the problem of the loading of classical data by introducing a quantum data-encoding scheme that provides control over both the number of qubits and the quantum circuit depth. We achieve this in two steps. First, we exploit the mapping between matrix-product states and quantum circuits to map each image into an MPS. We control the depth of the corresponding circuit via the bond dimension of the MPS, and we control the number of qubits by splitting the image into patches (where each patch is encoded as an independent MPS). We test this encoding by using an MPS-based classifier on the Fashion-MNIST dataset. This MPS-based approach can already then be directly implemented on a quantum computer. Second, we propose a hardware-efficient quantum circuit compression, which similarly allows for control over both the number of qubits and the circuit depth. In this case, however, the compression method is not limited by the entanglement of the quantum state in the same way as MPS. We demonstrate that a hardware-efficient quantum circuit classifier can achieve competitive accuracy on the Fashion-MNIST dataset using only 11 qubits. These two efforts together provide a scalable method to tune classification accuracy on quantum devices according to available hardware.

\section{Image classification using matrix-product states}
\label{sec:mps_classifier}
In this section, we describe the MPS approach for machine learning, including the data-encoding scheme and the classifier. We focus on the task of image classification on the Fashion-MNIST dataset, which contains 60000 training images and 10000 test images from ten label classes. In our experiments, we resize the default Fashion-MNIST images using a bilinear interpolation from $28\times 28$ to $32\times 32$ to facilitate the patching procedure that we introduce in the following section.

\subsection{Data encoding}\label{sec:encoding}
\begin{figure}
    \centering
    \includegraphics{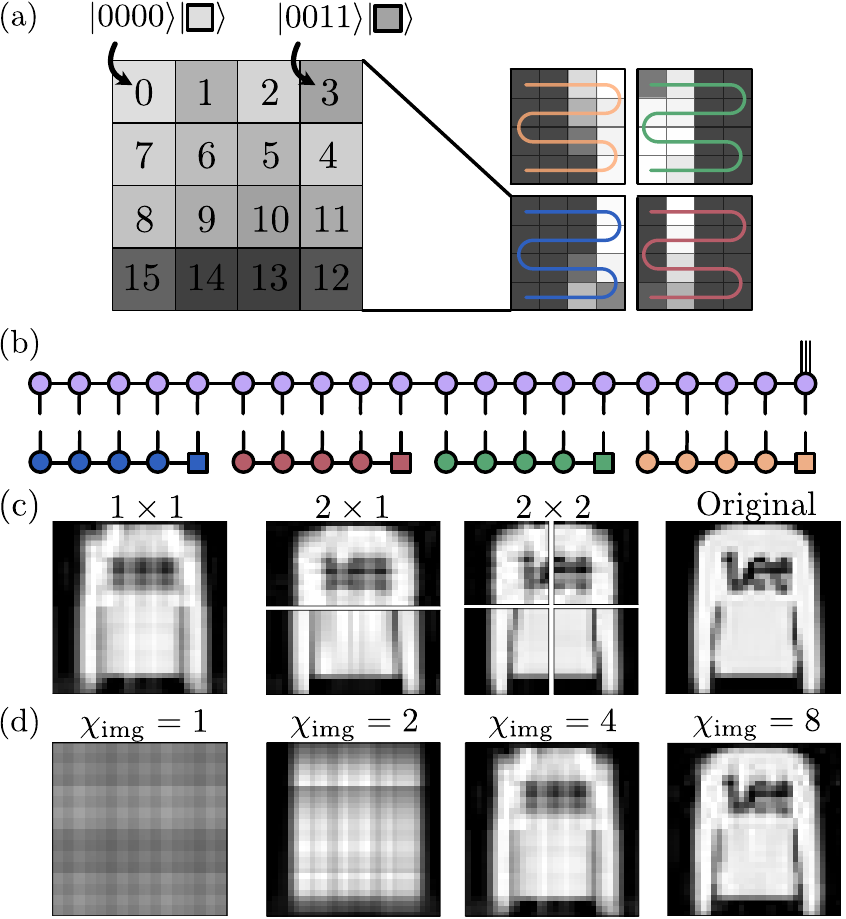}
    \caption{(a) Our encoding scheme consists of splitting the image into patches, each of which is encoded in a quantum superposition state consisting of address qubits and a color qubit, as defined in Eq.~\eqref{eq:FRQI}. (b) In the MPS learning protocol, each patch is separately encoded into an MPS consisting of tensors with physical dimension 2 corresponding to address (circle) qubits and the color (square) qubit. These MPS are concatenated and contracted with an MPS-based classifier (purple) with fixed bond dimension. An additional dimension 10 classifer leg, represented by three lines, is used to classify the Fashion-MNIST images. (c) An example encoded Fashion-MNIST image with bond dimension $\chi_{\text{img}}=4$ for different numbers of patches, compared with the original uncompressed image. (d) The compressed image as a single single patch with varying bond dimension $\chi_{\text{img}}=1,2,4,8$.}
    \label{fig:patching_schematic}.
\end{figure}

A standard way to encode classical images in a quantum system is the so-called flexible representation of quantum images (FRQI)~\cite{Le2010,Yan2016}, in which $N$ pixels are encoded using $\log_2{N}+1$ qubits. Each $N$-pixel grayscale image is viewed as a flattened $N$ dimensional vector $(p_0,\cdots, p_{N-1})$ with pixel values $p_i\in [0,1]$. This vector is then encoded to the following quantum state.
\begin{equation}\label{eq:FRQI}
    \vert\psi\rangle = \frac{1}{\sqrt{N}}\sum_{x=0}^{N-1}\vert x\rangle \left(\cos{\frac{\pi p_x}{2}}\vert 0\rangle + \sin{\frac{\pi p_x}{2}}\vert 1\rangle\right)
\end{equation}
The first $\log_2{N}$ qubits, which we refer to as \emph{address} qubits, label the pixel locations, i.e., the computational basis states $\vert x\rangle$ correspond to binary representations of the location in the $N$-dimensional array, see Fig.~\ref{fig:patching_schematic}a for a schematic. The remaining qubit, which we refer to as the \emph{color} qubit, encodes the pixel value or brightness. This encoding is similar to amplitude encoding~\cite{Schuld2018,latorre2005,ashhab2022quantum}, but the use of the color qubit allows for an absolute intensity scale for the image which is lost due to normalization of the state in amplitude encoding. Several quantum image processing algorithms that exhibit quantum speed-ups also rely on FRQI~\cite{schu2003,Zhang2015}.
By convention, we enumerate the pixels following a snake pattern as depicted in Fig.~\ref{fig:patching_schematic}a.

The FRQI uses quantum entanglement between the address and color qubits. For a generic image, this will require a circuit depth polynomial in $N$~\cite{Le2010, lin2021real}. Although FRQI uses only $\log_2N+1$ qubits to encode an $N$-pixel image, the hardware requirements shift from the qubits to the gates. To address the circuit depth, we instead use an approximate compressed representation based on matrix-product states (see App.~\ref{sm:sec:mps} for a brief review on MPS). The bond dimension, $\chi$, of the MPS limits the quantum entanglement and thus controls the accuracy of the approximation, as illustrated in Fig.~\ref{fig:patching_schematic}c. Importantly, there exists a direct mapping between MPS and sequential quantum circuits~\cite{Schon2005a, Smith2019b, ran2020encoding, Barratt2020, lin2021real}, as outlined in App.~\ref{sm:sec:mps}. The circuit depth scales linearly in the number of qubits and polynomially with $\chi$. The FRQI can thus be coupled with the MPS approximation to reduce the circuit depth to  $\mathcal{O}(\text{poly}(\chi) \log_2{N})$. 

To control the number of qubits, we can divide our image into patches and encode each patch independently using the FRQI (see Fig.~\ref{fig:patching_schematic}). If we split the image into $N_p$ patches, the encoding scheme requires $\left(\left \lceil{\log_2{ N/N_p}}\right\rceil + 1\right) N_p$ qubits. Taking $N_p = N$ means each pixel is encoded in a single qubit, as considered in Ref.~\cite{salvador2003,stoudenmire2016supervised}. The encoded image is a product state of $N$ qubits in the states $\vert\psi_x\rangle = \cos{\frac{\pi p_x}{2}}\vert 0\rangle + \sin{\frac{\pi p_x}{2}}\vert 1\rangle$. We refer to this as the single-pixel limit. This patching procedure allows us to interpolate between the FRQI limit and the single-pixel limit.

\subsection{The MPS classifier}\label{sec: MPS classifier}
\begin{figure}[!th]
    \centering
    \includegraphics[width=.45\textwidth]{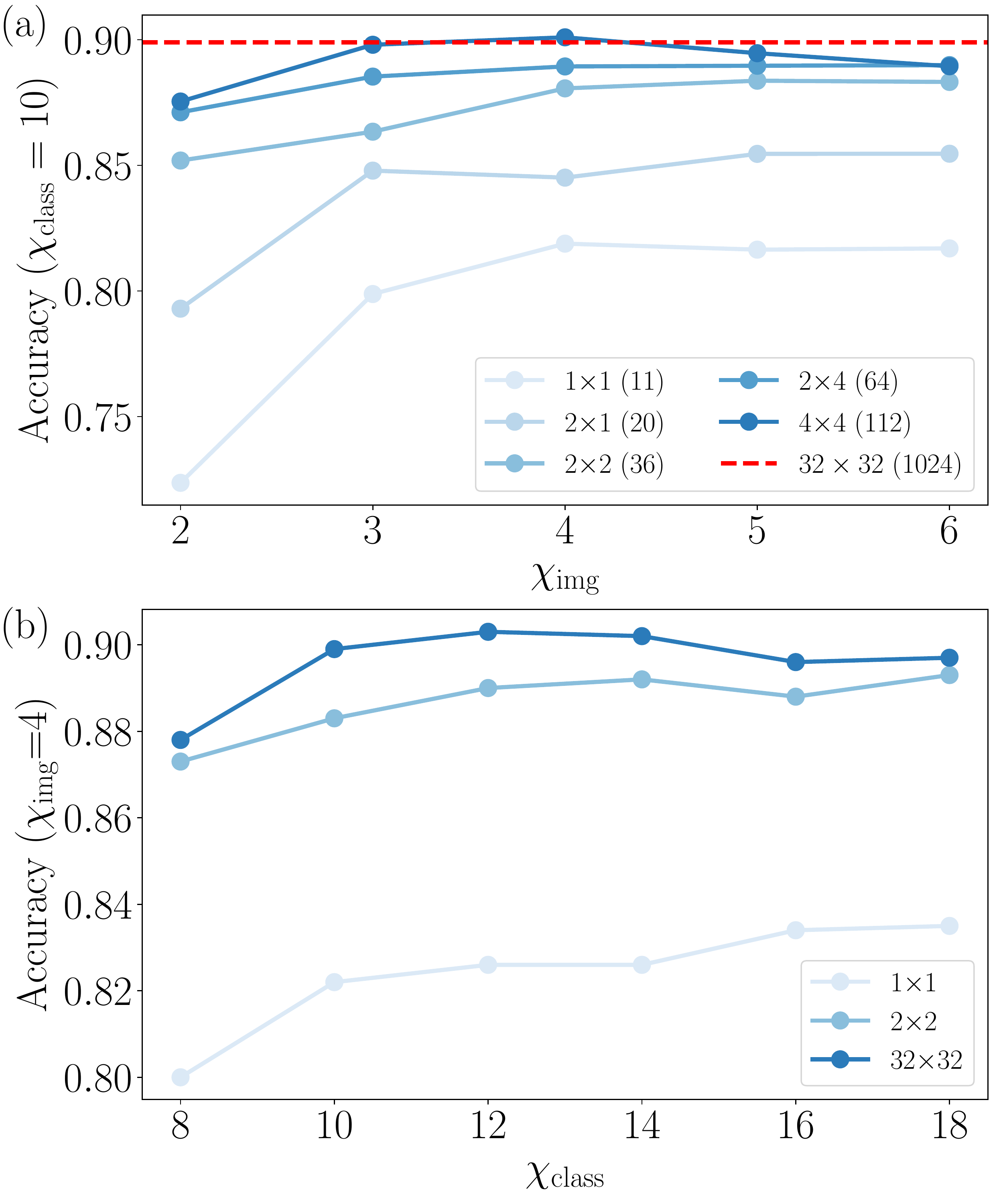}
    \caption{
    (a) The test accuracy for classification using MPS as a function of the image bond dimensions $\chi_\text{img}$ for fixed classifier bond dimension $\chi_\text{class}=10$. Each curve corresponds to a different number of patches, see Fig.~\ref{fig:patching_schematic}. We report the average of the best 100 test accuracies to limit stochastic effects. The numbers in parentheses in the legend are the required number of qubits on a quantum computer. The dashed red line shows the single pixel limit, which is an exact encoding of the image with $\chi_\text{img}=1$. (b) The test accuracy as a function of the classifier bond dimension $\chi_\text{class}$, for fixed image bond dimension $\chi_\text{img}=4$. 
    }
    \label{fig:fn_learning_results}
\end{figure}

To classify the different images, we train an MPS classifier~\cite{stoudenmire2016supervised,efthymiou2019tensornetwork} with dimension $2$ physical legs and a single additional dimension $L$ ``label'' leg, where $L$ is the number of labels ($L=10$ for Fashion-MNIST). We contract each image MPS with the classifier MPS; the element with largest amplitude in the resulting length $L$ vector is the predicted label. This contraction method is shown in Fig.~\ref{fig:patching_schematic}b. In our experiments we use the Adam optimizer~\cite{kingma2014adam} with learning rate $10^{-4}$ and batch size $128$ -- see App.~\ref{sm:mps_training}.

In Fig.~\ref{fig:fn_learning_results}a we show the test accuracy obtained when using our MPS data compression for various numbers of patches and bond dimensions $\chi_\text{img}$, where we fixed the classifier bond dimension $\chi_\text{class}=10$. We achieve performance comparable with state-of-the-art tensor network methods, but at a fraction of the hardware requirements. In particular, Ref.~\cite{efthymiou2019tensornetwork} achieved test accuracy of approximately 88\% on the Fashion-MNIST dataset by assigning a single qubit to each pixel, which would cost $784$ qubits using the original $28\times 28$ images. As shown in Fig.~\ref{fig:fn_learning_results}, we can achieve similar accuracy with relatively shallow circuits (i.e., bond dimension $2-3$) and only $64$ qubits, corresponding to the $2\times 4$ patch case. 

Additionally, we find that increasing both the bond dimensions $\chi_\text{img}$ (number of gates) and the number of patches (number of qubits) improves the test accuracy.

Notably, the accuracy as a function of the image bond dimension plateaus at a different point for each number of patches. This suggests that the number of patches (and number of qubits) is important for improving the accuracy. Since our method allows us to tune both parameters, it allows us to find an optimal compression of the image that respects the limitation of the device.

Fig.~\ref{fig:fn_learning_results}b shows the dependence of the test accuracy on the classifier bond dimension $\chi_{\text{class}}$, for a fixed $\chi_\text{img}$. We find that beyond $\chi_\text{class}=10$ increasing the bond dimension has a relatively small impact on the classification accuracy for most choices of patching. We similarly observe that increasing the number of patches increases the accuracy in all cases.

\section{Classification using quantum circuits}
In this section, we describe an approach to quantum machine learning based on parameterized quantum circuits~\cite{schuld2020,Benedetti2019,reza2021}. We use sequential circuits to both encode the classical data and to implement the classifier. The sequential circuit structure is inspired by the preceding MPS approach but is specifically tailored for the local and pairwise connectivity of many quantum computer realizations, and so we refer to them as hardware-efficient.

\subsection{Quantum data encoding}
\label{sec:data_load_qc}
\begin{figure}[t]
    \centering
    \includegraphics{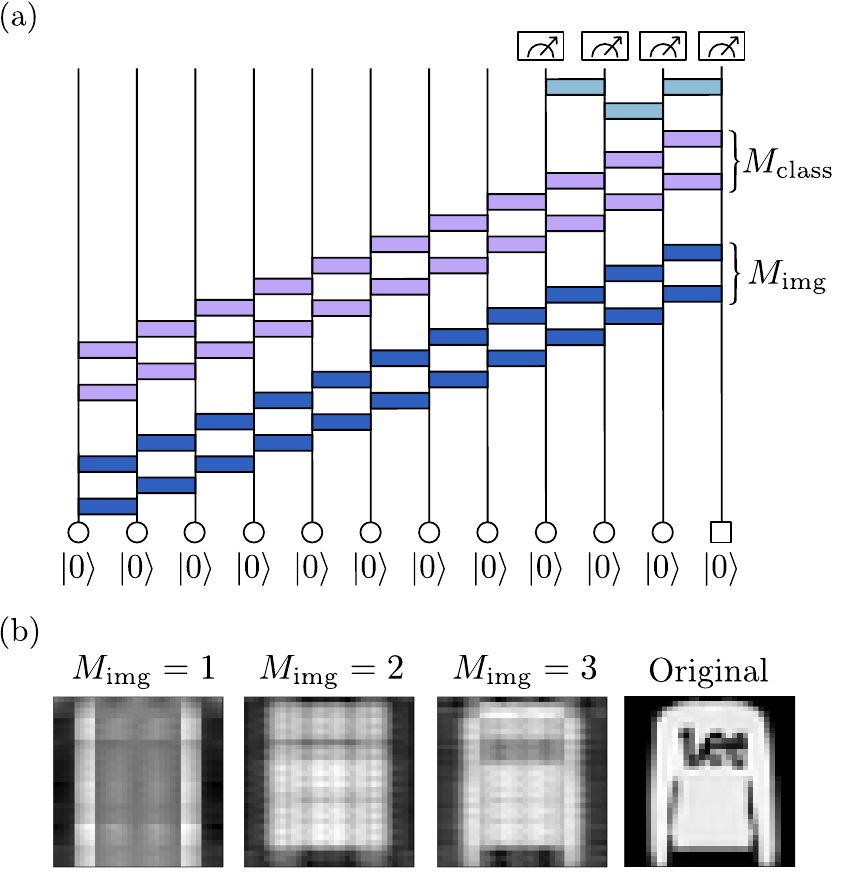}
    \caption{
    (a) The quantum circuit encoding and classification scheme using 11 qubits. We compress each image using $M_{\text{img}}$ layers of sequential circuits consisting of two-qubit gates. The address and color qubits are indicated by circles, and a square, respectively, similar to Fig.~\ref{fig:patching_schematic}. We classify each image with $M_{\text{class}}$ layers of sequential two-qubit gates (the case $M_{\text{img}} = M_{\text{img}} = 2$ is shown for illustration). Before the final readout, three additional trainable two-qubit gates are added. (b) Compression of a selected image using $M_{\text{img}} = 1,2,3$, compared with the uncompressed image.}
    \label{fig:circuit_compress}
\end{figure}

When mapping the MPS based approach of the previous section to a quantum circuit, we are left with a circuit depth that scales polynomially with $\chi$. This is because the mapping entails a sequence of multi-qubit gates where each gate acts on $\lceil\log_2\chi\rceil+1$ qubits, which must subsequently be decomposed into the two-qubit gates and rotations implemented on physical devices (see App.~\ref{sm:sec:mps}). We propose an alternative circuit structure for encoding the classical data as a quantum state, which consists of $M_\text{img}$ layers of sequentially arranged two-qubit gates, as shown in Fig.~\ref{fig:circuit_compress}a. These gates are parameterized and optimized such that the resulting state has maximal fidelity with the exact encoding of the state. Note that since the pixel values are implicitly contained in the probabilities for measuring each of the computational basis states, the optimization can in principle be performed without computing overlaps of states on the quantum computer. We open-source these processed circuits at \href{https://zenodo.org/record/6562229}{https://zenodo.org/record/6562229}. 

In Fig.~\ref{fig:circuit_compress}b, we display a sample scaled $32\times 32$ image compressed using the sequential ansatz for $M_{\text{img}} = 1,2$ and $3$, using the FRQI encoding on $11$ qubits. We can additionally include the patching procedure to control the number of qubits used. However, due to the computational cost of simulating the quantum circuits, we restrict ourselves to a single patch $N_p=1$. We use the Adam optimizer to obtain the optimal circuit compression.

The sequential circuit structure that we use is a subclass of MPS with bond dimension $\chi = 2^{M_\text{img}}$~\cite{Schon2005a, Smith2019b, ran2020encoding, Barratt2020, lin2021real}, as explained in App.~\ref{sm:sec:mps}. To generate entanglement entropy $S \sim \log\chi$ requires exponentially fewer parameters in our quantum circuit. Conversely, for the same number of parameters, our quantum circuits generate more entanglement.

\subsection{Quantum circuit classifier}
To classify the encoded images, we similarly use a hardware-efficient sequential circuit with $M_\text{class}$ layers, as shown in Fig.~\ref{fig:circuit_compress}a. It is possible to directly implement the MPS classifier in Sec.~\ref{sec: MPS classifier} as a quantum circuit but with two undesirable features. The first is that, similarly to the state, the circuit will consist of multi-qubit gates set by the bond dimension. The second is that this approach requires projections for some of the qubits. The result is that the number of shots (runs of the circuits) required to accurately measure the classification outcome scales exponentially with the number of qubits used. 

To classify the images we measure the four right-most qubits in Fig.~\ref{fig:circuit_compress}a. Of the $16=2^4$ bit string outcomes, the first $10$ correspond to the classes for our images. The classification is made by taking the bit string with highest probability. Note that following our sequential layers we include three additional gates before measuring, as shown in Fig.~\ref{fig:circuit_compress}a. These ensure that information can propagate from the bottom color qubit to all measured qubits. For $M_\text{class}\geq 3$ these extra gates are not required but improve the accuracy of the classification.

\begin{figure}
    \centering
    \includegraphics[width=.45\textwidth]{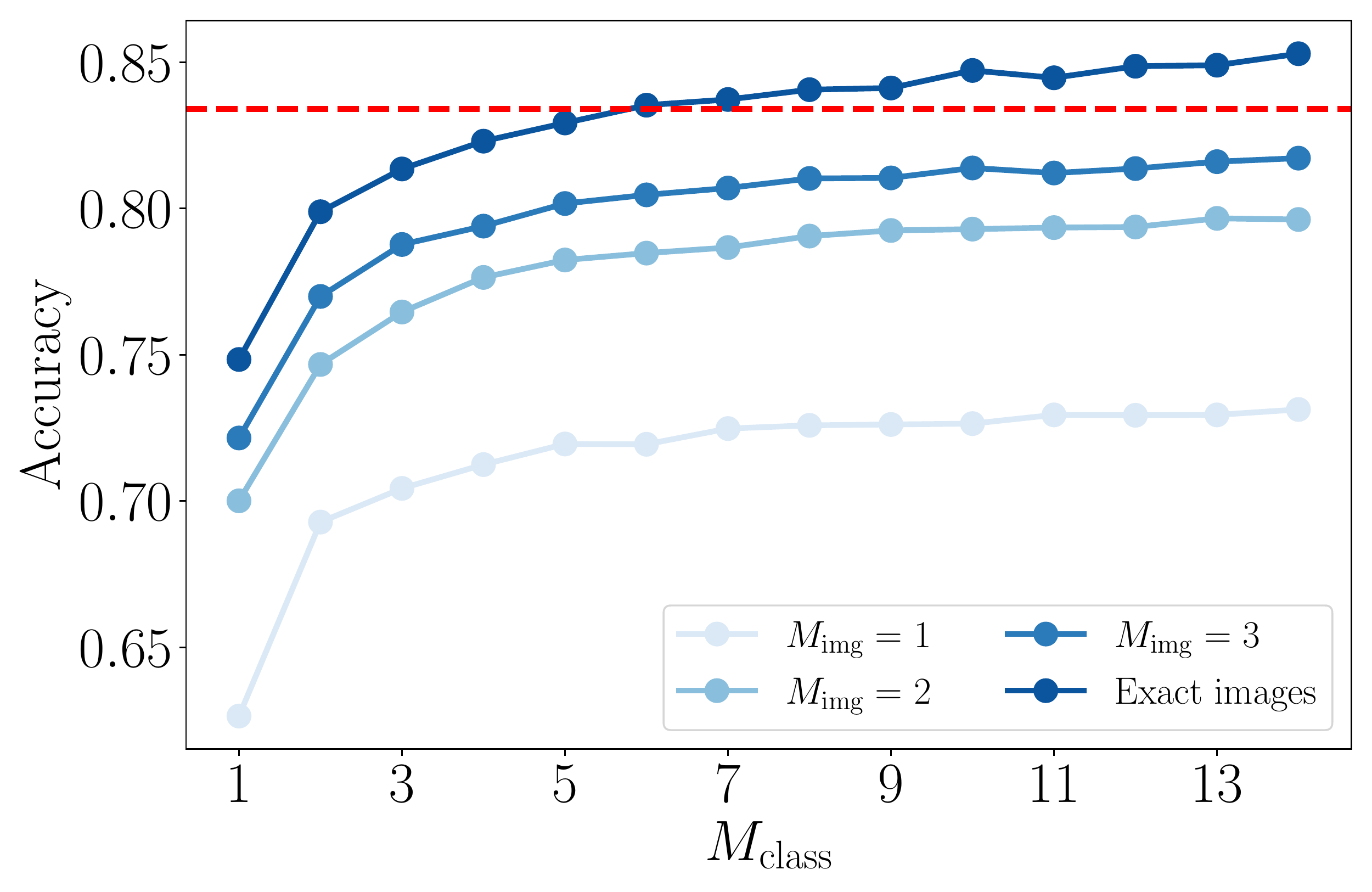}
    \caption{
    The test accuracy using hardware-efficient quantum circuits for image compression and classifier on the Fashion-MNIST dataset. We again report the average of the best $100$ converged iterations to limit stochastic effects. The accuracy is plotted against the number of layers $M_{\text{class}}$ in the quantum circuit classifier. We show the results the images compressed using $M_{\text{img}} = 1, 2, 3$, compared with the exact uncompressed images. For reference, we show the accuracy we achieve for the $\chi_{\mathrm{img}}=4$ single patch MPS case with $\chi_{\mathrm{class}}=16$.
    }
    \label{fig:test_qc}
\end{figure}

We report the test accuracy achieved on the Fashion-MNIST dataset using our quantum circuit approach in Fig.~\ref{fig:test_qc}. As we increase the number of layers in the encoded image state $M_\text{img}$, we see a significant increase in the classification accuracy. We additionally include the results for the exact state encoded using FRQI, which are quickly approached by increasing the layers in the image encoding. Moreover, as a function of $M_\text{class}$ the accuracy appears to plateau for small values. This shows that with only a modest number of layers in both the state and the classifier we can achieve competitive classification accuracy. This demonstrates that our method facilitates effective classification with resource requirements that are realistic for NISQ quantum computers.

We also note the dashed red line in Fig.~\ref{fig:test_qc}, which corresponds to an MPS experiment with $\chi_{\mathrm{class}}=16$ and $\chi_{\mathrm{img}}=4$. The circuit contains far fewer parameters but nonetheless achieves competitive accuracy for the $M_{\text{img}}=3$ case.

\section{Discussion}\label{sec:discussion}
In this paper, we proposed encoding and compression schemes for processing classical data on NISQ devices. Our approach provides the control over the required physical resources, namely the number of qubits and the circuit depth. Furthermore, we demonstrated that using hardware-efficient circuits for both the data encoding and classifier, we can achieve competitive accuracy on the Fashion-MNIST dataset. Having established the capabilities of hardware-efficient circuits on image classification problems, the protocol we use in our MPS experiments provides a straightforward method to scale accuracy to hardware availability. 

We note that the investigation of patching for the quantum circuit case would be significantly more difficult on a classical computer; the clock-time required to compress the full dataset then subsequently optimize the highly entangled sequential circuit is prohibitive. For a small number of layers, MPS based methods could be used, but these also become infeasible as the number of layers increases. On the other hand, the patching can be efficiently implemented on a near-term quantum device.

The quantum circuits are shallow representations capable of efficiently encoding long range entanglement. We contrast the circuits with matrix product states, which are ideally suited to encode locally entangled states. It is interesting to consider whether quantum advantages can be achieved exploiting  different ways of encoding entanglement, depending on the learning task and dataset. 

A natural extension of our work is to consider various other circuit and MPS structures, such as brickwall-patterned circuits, MERA~\cite{vidal2007}, and higher dimensional variants. One could also consider hybrid architectures where neural networks act as autoencoders that preprocess the inputs to the quantum architecture. We also note that although the best image recognition methods on Fashion-MNIST typically achieve performances of ~96\%~\cite{tanveer2021fine, foret2020sharpness}, they require several million parameters, which we contrast with the several thousand that large MPS and hardware efficient circuits would require. 

Furthermore, while we discussed two methods for classification, our image compression scheme can be used more generally. Improvements to the quantum classifier, for instance by incorporating additional structure or matching the connectivity of NISQ devices, remain interesting open questions. Additionally, as with any quantum optimization problem, a realistic algorithm should take into account the effects of gate errors and decoherence. Nevertheless, the approach we introduce allows for practical machine learning tasks to be performed with realistic quantum resources, requiring as few as 11 qubits. The encoded Fashion-MNIST images can be used as a quantum dataset for benchmarking quantum classifiers. By providing control over the number of qubits and circuit depth, we have introduced a flexible image encoding approach for the NISQ-era and beyond.

\section{Acknowledgements}
R.D. acknowledges ShengHsuan Lin for helpful discussions and technical assistance. Y.-J.L was supported by the Max Planck Gesellschaft (MPG) through the International Max Planck Research School for Quantum Science and Technology (IMPRS-QST). A.S. was partly supported by a Research Fellowship from the Royal Commission for the Exhibition of 1851. F.P. acknowledges support of the European Research Council (ERC) under the European Unions Horizon 2020 research and innovation program (grant agreement No. 771537). F.P. also acknowledges the support of the Deutsche Forschungsgemeinschaft (DFG, German Research Foundation) under Germany’s Excellence Strategy EXC-2111-390814868. F.P.’s research is part of the Munich Quantum Valley, which is supported by the Bavarian state government with funds from the Hightech Agenda Bayern Plus.

\bibliographystyle{apsrev4-2}
\bibliography{ref_editable}
%
%
\appendix
\section{Matrix-product states}
\label{sm:sec:mps}
\begin{figure}
    \centering
    \includegraphics{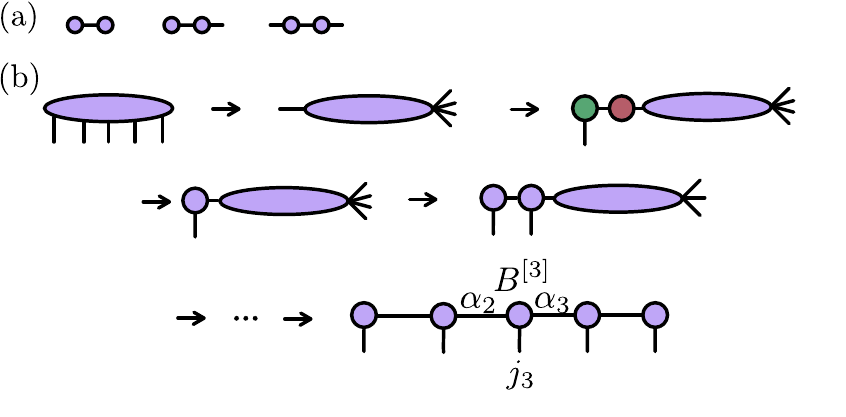}
    \caption{(a) Basic MPS diagrammatic notation. Connected tensors represents summing over the index corresponding to the connected legs. From left to right, a vector inner product, a matrix-vector multiplication, and a matrix-matrix multipication. (b) Converting a general vector into an MPS. A generic vector with size $2^N$ can be viewed as an $N$ index tensor. We convert each vector into a product of matrices via a singular value decomposition, then truncate the singular value matrix (the red tensor) to the desired number of singular values (i.e., $\chi$). By repeatedley applying this procedure, we decompose the original vector into a matrix-product state with bond dimension $\chi$. To return to the vector representation, we would contract the MPS along the virtual indices $\alpha_j$ (albeit with some loss from the truncation).}
    \label{sm:fig:mps_basics}
\end{figure}

Matrix-product states are an ansatz class where the coefficients of a full $n$-qubit state $\psi$ are decomposed into products of matrices. Explicitly, 
\begin{equation}
    \label{eq:mps}
    \begin{split}
    \ket{\psi} = \sum\limits_{\{j_k\}}\sum\limits_{\{\alpha_l\}}&B^{[1]j_1}_{\alpha_1}B^{[2]j_2}_{\alpha_1\alpha_2}\dots B^{[n]j_n}_{\alpha_{n-1}}\ket{j_1,j_2,\dots,j_n},
    \end{split}
\end{equation}
where the $j_k\in\{0,1\}$ indices are referred to as ``physical'' indices and the $\alpha$ indices are referred to as ``virtual indices.'' By convention, we refer to the dimension of the $\alpha$ indices as the bond dimension $\chi$. Without loss of generality, we may assign any MPS a single bond dimension $\chi$ corresponding to the largest bond dimension in the network (padding the others with zeros). In the above expression, each tensor $B^{[k]}$ for $k=2,\dots,n-1$ contains three indices, two virtual and one physical. The boundary tensors $B^{[1]}$ and $B^{[n]}$ each contain two indices. We will often use a diagrammatic notation for tensor manipulations, where tensors are represented by symbols and each index is represented by a leg. Two connected legs represent summing over the corresponding index. See Fig.~\ref{sm:fig:mps_basics}a for more details.

A length $2^n$ vector can always be decomposed into a matrix-product state with $n$ tensors; see Fig.~\ref{sm:fig:mps_basics}b. Decomposing a vector in this way will guarantee that the tensors satisfy the isometry condition

\begin{equation}
    \label{eq:isometry}
    \sum_{j_k}\sum\limits_{\alpha_k} B^{[k]j_k}_{\alpha_{k-1}\alpha_k}     \left(B^{[k]j_k}_{\alpha'_{k-1}\alpha_k}\right)^* = \delta_{\alpha_{k-1},\alpha'_{k-1}}.
\end{equation}

\begin{figure}
    \centering
    \includegraphics{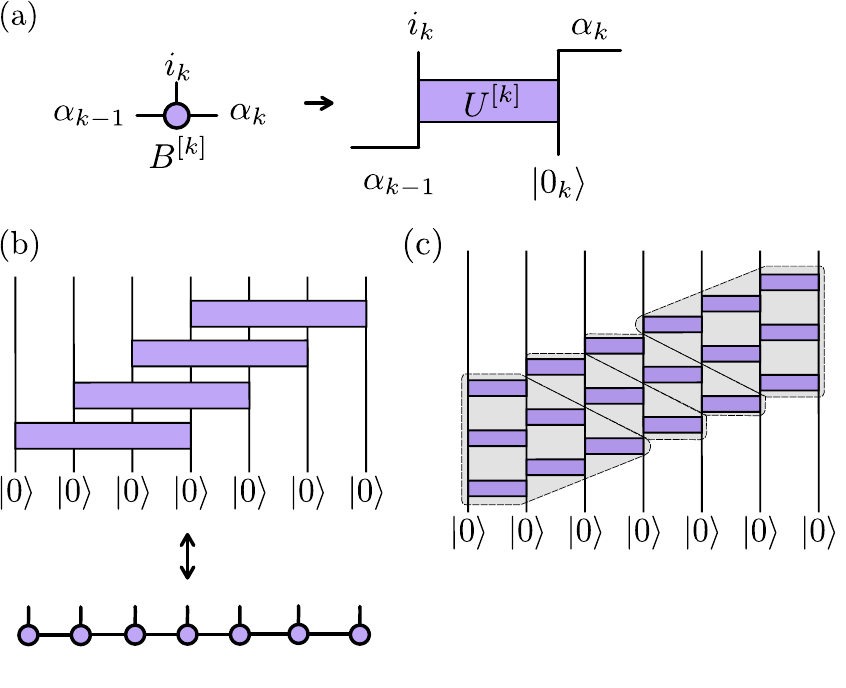}
    \caption{
    (a) The mapping between an MPS tensor and a unitary matrix. The MPS tensors are used to define a unitary action on an additional set of qubits initialized to a product state, where the number of additional qubits depends on the bond dimension of the MPS. (b) An MPS with bond dimension $\chi$ is equivalent to a quantum circuit where each gate acts on $\log{\chi} + 1$ qubits. (c) Every $M$-layer sequential quantum circuit with two qubit gates is a subset of the set of sequential quantum circuit with $(M+1)$ qubit gates. Here, we show the equivalence for $M=3$.
    }
    \label{sm:fig:mps_qc}
\end{figure}

An MPS with bond dimension $\chi$ where all the tensors satisfy Eq.~\ref{eq:isometry} can be exactly mapped to a sequential quantum circuit with unitaries acting on $\log\chi + 1$ qubits, as shown in Fig.~\ref{sm:fig:mps_qc}. For practical implementations, each unitary gate must be further decomposed into single and two-qubit gates. For a generic quantum gate acting on $\log_2\chi+1$ qubits, this requires $O(\text{poly}(\chi))$ single and two-qubit gates~\cite{qc_book}, resulting in a total cost of $O(\text{poly}(\chi)n)$ quantum operations.

The mapping, which is diagrammatically depicted in Fig~\ref{sm:fig:mps_qc}a, is given by
\begin{equation}
    B^{[k]j_k}_{\alpha_{k-1}\alpha_k} = \langle\alpha_k, j_k\vert U^{[k]}\vert 0_k,\alpha_{k-1}\rangle,
\end{equation}
where $\ket{0_k}$ is a product state. We refer the reader to~\cite{lin2021real} for more details.

On the other hand, a sequential quantum circuit with $M$ layers of two-qubit gates can be viewed as an equivalent sequential circuit with a single layer of $M+1$ qubit gates (see Fig~\ref{sm:fig:mps_qc}b). This circuit, in turn, can be mapped to an MPS with bond dimension $\chi=2^M$. Every single-layer circuit thus has an exact $\chi=2$ equivalence.

\section{MPS training}
\label{sm:mps_training}
\begin{figure}
    \centering
    \label{fig:mps_training}
    \includegraphics[width=.45\textwidth]{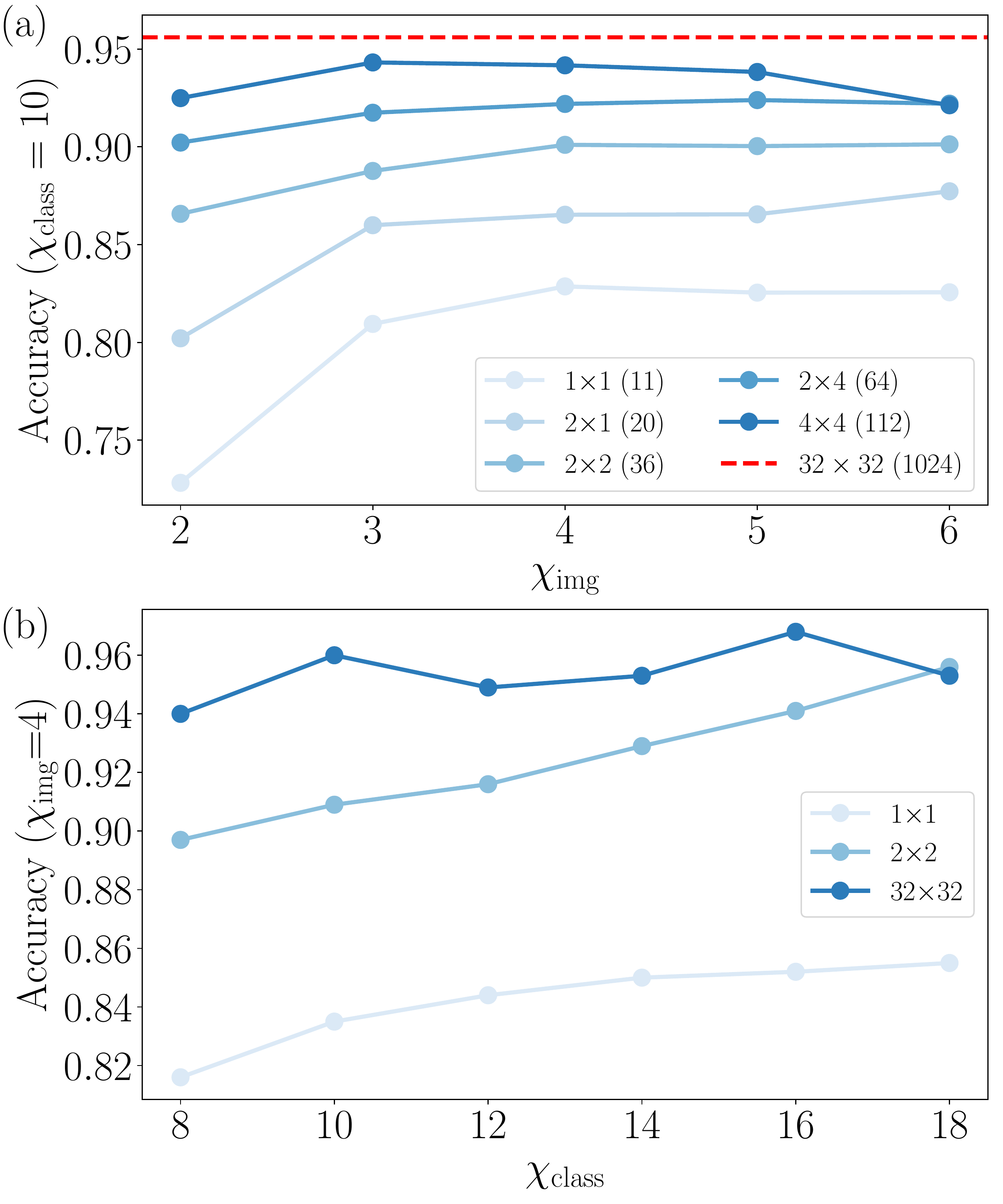}
    \caption{(a) The training accuracy as a function of the image bond dimension. (b) The accuracy as a function of classifier bond dimension.}
\end{figure}

As discussed in the main text, we train using the Adam optimizer with learning rate $10^{-4}$ and a minibatch size of $N_b=128$. We trained for $3000$ epochs for most cases (convergence generally occurred well before this). In Figure~\ref{fig:mps_training}, we show the training accuracy as a function of the bond dimension and patch size.

For training, we made use of the Jax library~\cite{jax2018github}. We used Pytorch ~\cite{paszke2019pytorch} to load and transform the datasets. For training, we used a log softmax cross entropy loss function with $L_2$ regularization. 

Given the classifier output vectors and their corresponding labels within a minibatch (together denoted by $\{(\bm{s},t)\}$), the loss function is defined by

\begin{equation}
    \label{sm:eq:loss}
    L(\{(\bm{s},t)\}) = -\frac{1}{N_b}\sum_{(\bm{s}, t)}\log\bigg(\frac{e^{Cs_t}}{\sum\limits_j e^{Cs_j}}\bigg) + \frac{\lambda}{2N_b}\sum\limits_{w_i} w_i^2,
\end{equation}

where the sum is over all tuples of classifier output $\bm{s}$ and the correct label $t$ for the corresponding image. In the above equation, $s_j$ is the $j^\mathrm{th}$ element of vector $\bm{s}$, $w_i$ are the weights in the classifier MPS, $C$ is a constant used to avoid vanishing gradients, and $N_b$ is the minibatch size. In our experiments, we set $\lambda=10^{-4}$ and $C=1$. 

We initialized our classifier MPS using stacked identity matrices with Gaussian noise centered at $0$ with width $10^{-4}$. For the most part, the choice of initialization had minimal impact on the training, but the random noise needed to be sufficiently small to prevent exploding loss functions stemming from exponential buildup due to the sequential nature of a tensor network. This was occasionally an issue in training as well; we resolved it by factoring out the norm of the tensor network as needed, since we ultimately only cared about the relative values of the output prediction vector. The training accuracy is shown in Fig.~\ref{fig:mps_training}. We note that the plot is not monotonic. The classifier MPS reaches a point of maximal accuracy. For short matrix-product states, this leads to overtraining. For longer matrix-product states, this leads to some degree of overtraining, but eventually the variations in the Adam optimizer build until the output explodes (again because of the sequential nature of an MPS, a small change in the tensors will build exponentially). The degree of overtraining in the longer matrix-product states is thus somewhat variable; we note that the test accuracy in Fig~\ref{fig:fn_learning_results} is much cleaner, and ultimately is the important property.

\section{Quantum circuit training}

Our quantum circuit classifiers are sequential circuits of two-qubit gates, where each gate is a $4\times 4$ unitary parameterized by 15 parameters as $\exp\left(-\frac{i}{2}\sum_{\rho,\gamma\in\{0,1,2,3\}}\theta_{\rho,\gamma}\hat\sigma^{\rho}\otimes\hat\sigma^{\gamma}\right)$, with the matrices $\hat\sigma^0 = \mathbf{I},\ \hat\sigma^1 =\hat\sigma^x,\ \hat\sigma^2=\hat\sigma^y$ and $\hat\sigma^3=\hat\sigma^z$. We set $\theta_{0,0} = 0$ to fix the phase degree of freedom of the gate.

An input image state $\ket{\psi}_{\mathrm{img}}$ is classified by feeding it to a circuit classifier $\hat{U}$, then measuring a subset of qubits $\mathcal{L}$, which we call the label qubits. We denote the remainder of qubits in the system by $\mathcal{S}$. The measurements yield a probability vector $\bm{s}$, with elements given by
\begin{equation}
    s_{i_\mathcal{L}} = \sum_{i_{\mathcal{S}}} |\langle i_{\mathcal{L}}i_{\mathcal{S}}|\hat U|\psi\rangle|^2,
\end{equation}

where $i_{\mathcal{S}}$ and $i_{\mathcal{L}}$ denote the bitstrings corresponding to the computational basis states for the respective qubit sets. Note that the probability is normalized, i.e., $\sum_{i_{\mathcal{L}}} s_{i_\mathcal{L}} = 1$. The prediction is given by $\argmax_{i_{\mathcal{L}}} s_{i_\mathcal{L}}$. Because Fashion-MNIST contains $10$ label classes, we use four label qubits. This outputs a $2^4=16$ length vector, and we disregard the final six bitstrings.

\begin{figure}[H]
    \centering
    \includegraphics[width=.45\textwidth]{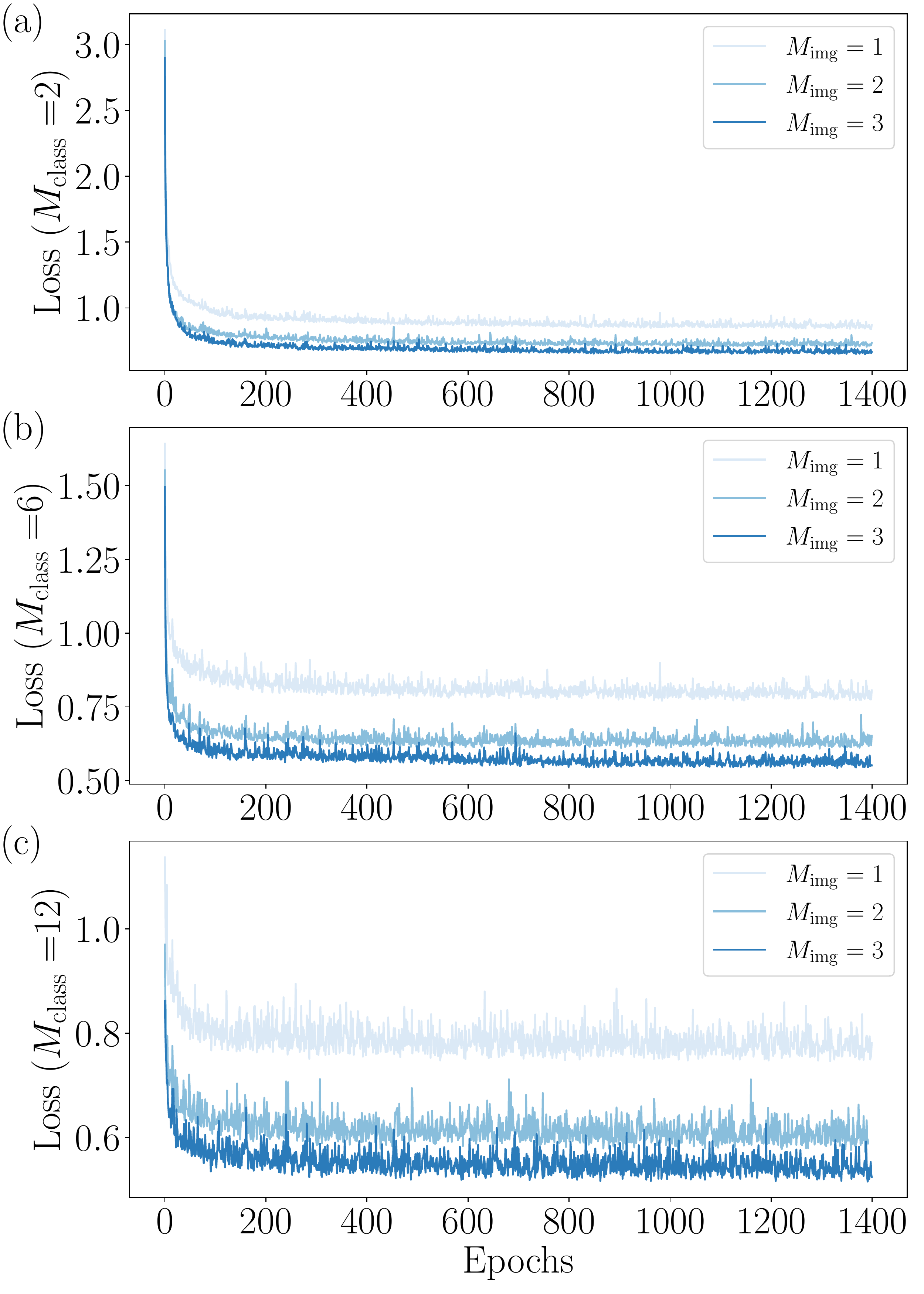}
    \includegraphics[width=.45\textwidth]{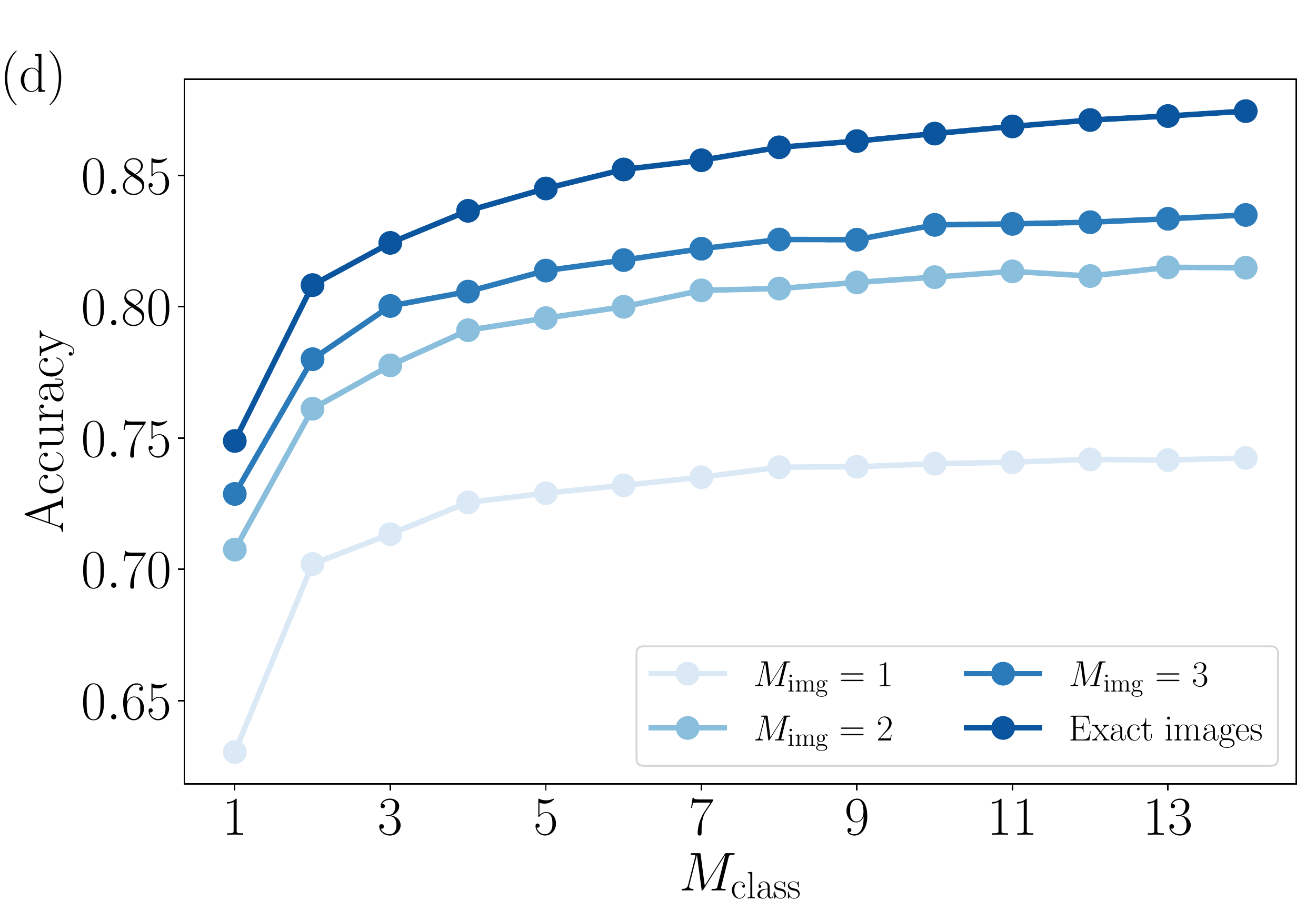}
    \caption{(a-c) Cross-entropy loss across training for the quantum circuit classifier. We show classifiers with $M_{\text{class}} = 2,6$ and $12$ respectively. In each case, we show $M_{\mathrm{img}} = 1,2,$ andA $3$. (d) Training accuracy for the quantum circuit classifier. Each curve corresponds to a different $M_{\mathrm{img}}$.}
    \label{fig:loss_acc_curve}
\end{figure}

In training our circuit model, we used the same loss function as in Eq.~\ref{sm:eq:loss}. We did not use any regularization ($\lambda=0$), and chose $C=N$ (the number of pixels). We use minibatch size $N_b=100$ for $1600$ epochs, and train using the Adam optimizer with learning rate $8\times10^{-4}$. The loss and training accuracy are displayed in Figure~\ref{fig:loss_acc_curve} for several values of $M_{\mathrm{class}}$.

\end{document}